\documentstyle[sprocl,cite,graphics,epsfig]{article}

\def\smallx{\scalebox{0.8}{SMALLX}}
\def\ldcmc{\scalebox{0.8}{LDCMC}}
\def\cascade{C\scalebox{0.8}{ASCADE}}

\newcommand{\SMALLX}{SMALLX}
\newcommand{\CCFM}{CCFM}

\newcommand{\alphasb}{\bar{\alpha}_s}

\def\be{\begin{equation}}
\def\ee{\end{equation}}
\def\bea{\begin{eqnarray}}
\def\eea{\end{eqnarray}}
\begin{document}

\vspace*{-3cm}
\begin{flushright}
  LU-TP 00-27\\
  hep-ph/0006166\\
  June 2000
\end{flushright}
\vspace*{1cm}

\title{Monte Carlo Generators and the CCFM Equation\footnote{Talk presented at the DIS 2000 workshop, Liverpool, England, April 2000.}}

\author{Leif L\"onnblad}

\address{Department of Theoretical Physics,\\
Lund University, Sweden\\E-mail: leif@thep.lu.se}

\author{Hannes Jung}

\address{Department of Physics,\\ Lund University, Sweden\\E-mail: jung@mail.desy.de} 


\maketitle\abstracts{We discuss three implementations of the CCFM
  evolution equations in event generator programs. We find that some
  of them are able to describe observables such as forward jet rates
  in DIS at HERA, but only if the so-called consistency constraint is
  removed. We also find that these results are sensitive to the
  treatment of non-singular terms in the gluon splitting function.}

\section{Introduction}

Small-$x$ final states are central to the understanding of small-$x$
evolution in general. To understand final-state properties such as
forward jet rates and transverse energy flow, it is important to have
event generators which give a good description of the experimental
data. It has, however, turned out to be difficult to produce such
an event generator.

To leading double-log accuracy, the CCFM\cite{CCFM} evolution should
be the best way of describing small-$x$ final states, and some
attempts to implement CCFM into an event generator has been made e.g.\ 
\smallx\cite{SMALLX} and \ldcmc\cite{ldcmc}. Below we will discuss
some recent developments of these generators as well as the new
\cascade\cite{cascade} generator for CCFM.

\section{CCFM and the SMALLX and CASCADE programs}

The implementation of the CCFM~\cite{\CCFM} parton evolution in the
forward evolution Monte Carlo program \smallx\ is described in detail
in~\cite{\SMALLX}, and we have already reported on some recent
developments in \cite{cascade} and \cite{Durham}. The main new
ingredient is a modification in the treatment of the non-Sudakov form
factor, $\Delta_{ns}$, which allows for the removal of the so-called
consistency constraint. Rather than using the simple form,
$\log\Delta_{ns} = -\alphasb\log(1/z_i)\log(k^2_{ti}/z_i q^2_{ti})$,
which requires the consistency constraint $k^2_{ti} > z_i q^2_{ti}$ in
order to be below unity, the full form\cite{Martin_Sutton} is used:
\begin{equation}
  \log\Delta_{ns} = -\alphasb
  \log\left(\frac{z_0}{z_i}\right)
  \log\left(\frac{k^2_{ti}}{z_0z_i q^2_{ti}}\right),
  z_0 = \left\{ \begin{array}{ll}
      1 & \mbox{if  } k_{ti}/q_{ti} > 1 \\
      k_{ti}/q_{ti} & \mbox{if  } z < k_{ti}/q_{ti} \leq 1 \\
      z             & \mbox{if  } k_{ti}/q_{ti} \leq z  
    \end{array} \right.
  \label{ns_new}
\end{equation} 
This form factor is well behaved for all emissions and gives no
suppression in the region $k_{ti}/q_{ti} \leq z$ where then
$\Delta_{ns}=1$.

\begin{figure}[t]
  \begin{center}
    \vspace*{-1cm}
    \hbox{
      \hspace*{-4.5cm}
      \vbox{\epsfig{figure=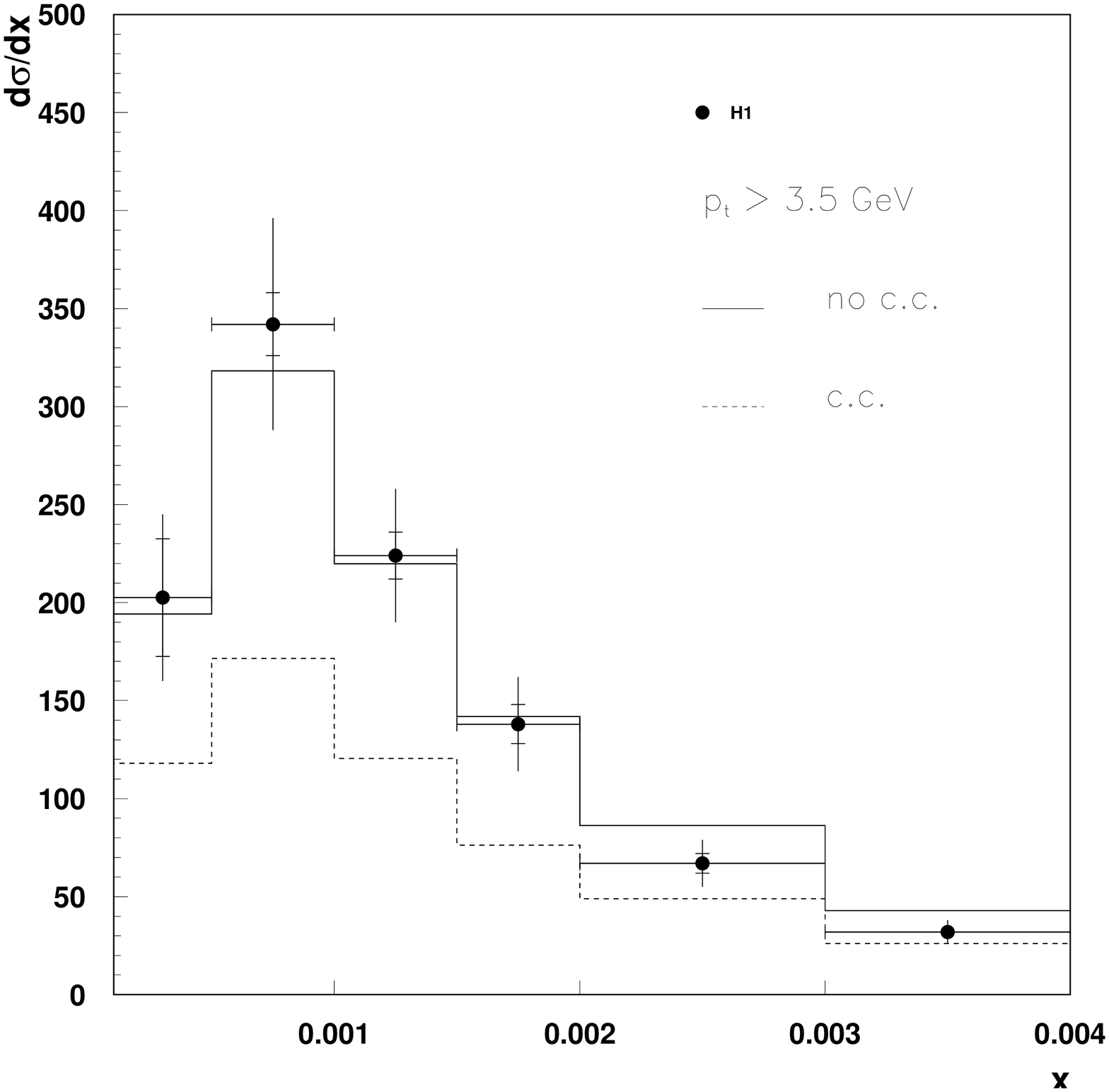,width=4cm,height=4.87cm}
        \vspace*{4pt}}
      \hspace*{-8.0cm}
      \vbox{\epsfig{figure=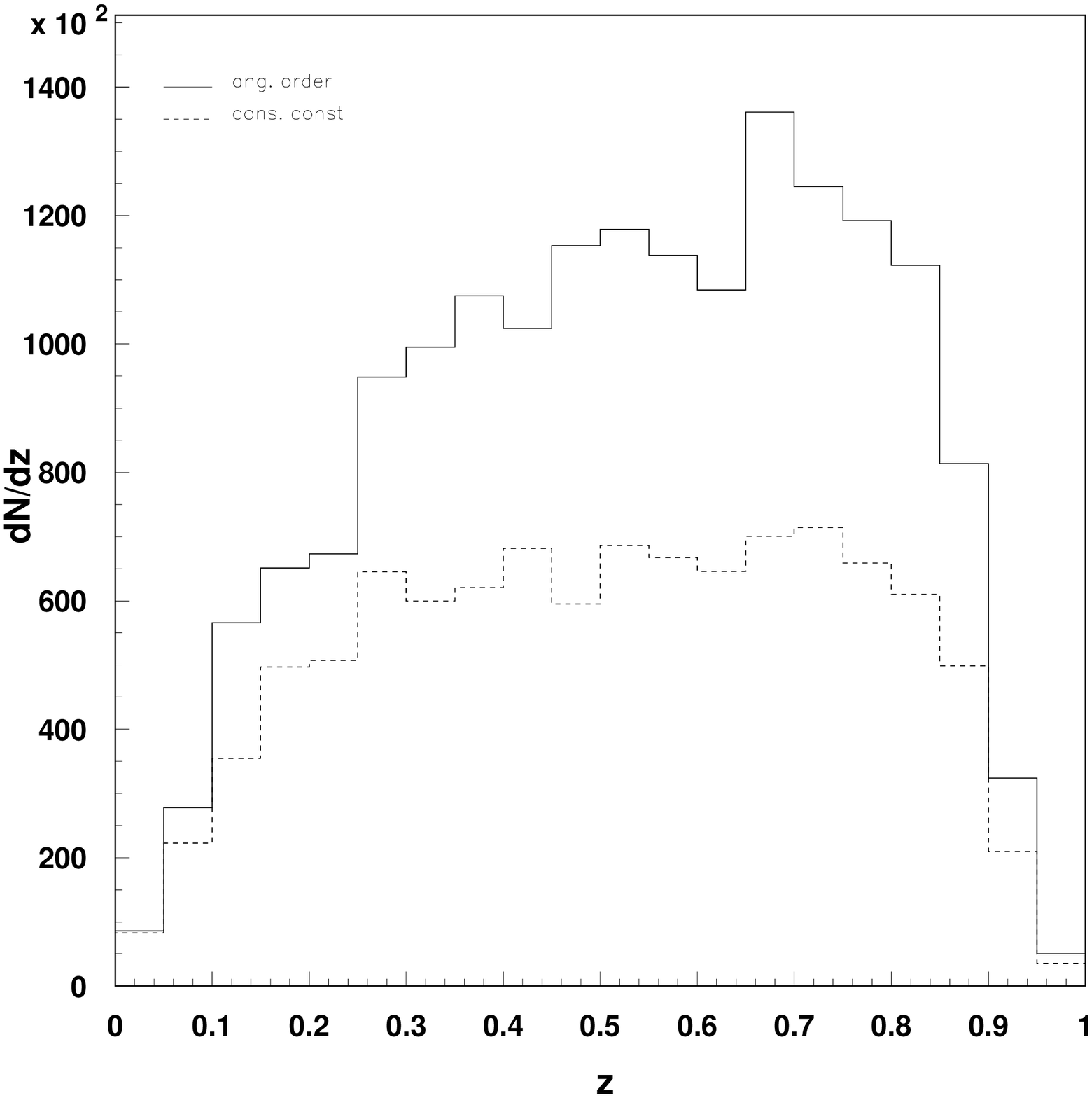,width=4cm,height=5cm}}
      \hspace*{-8.0cm}
      \vbox{\epsfig{figure=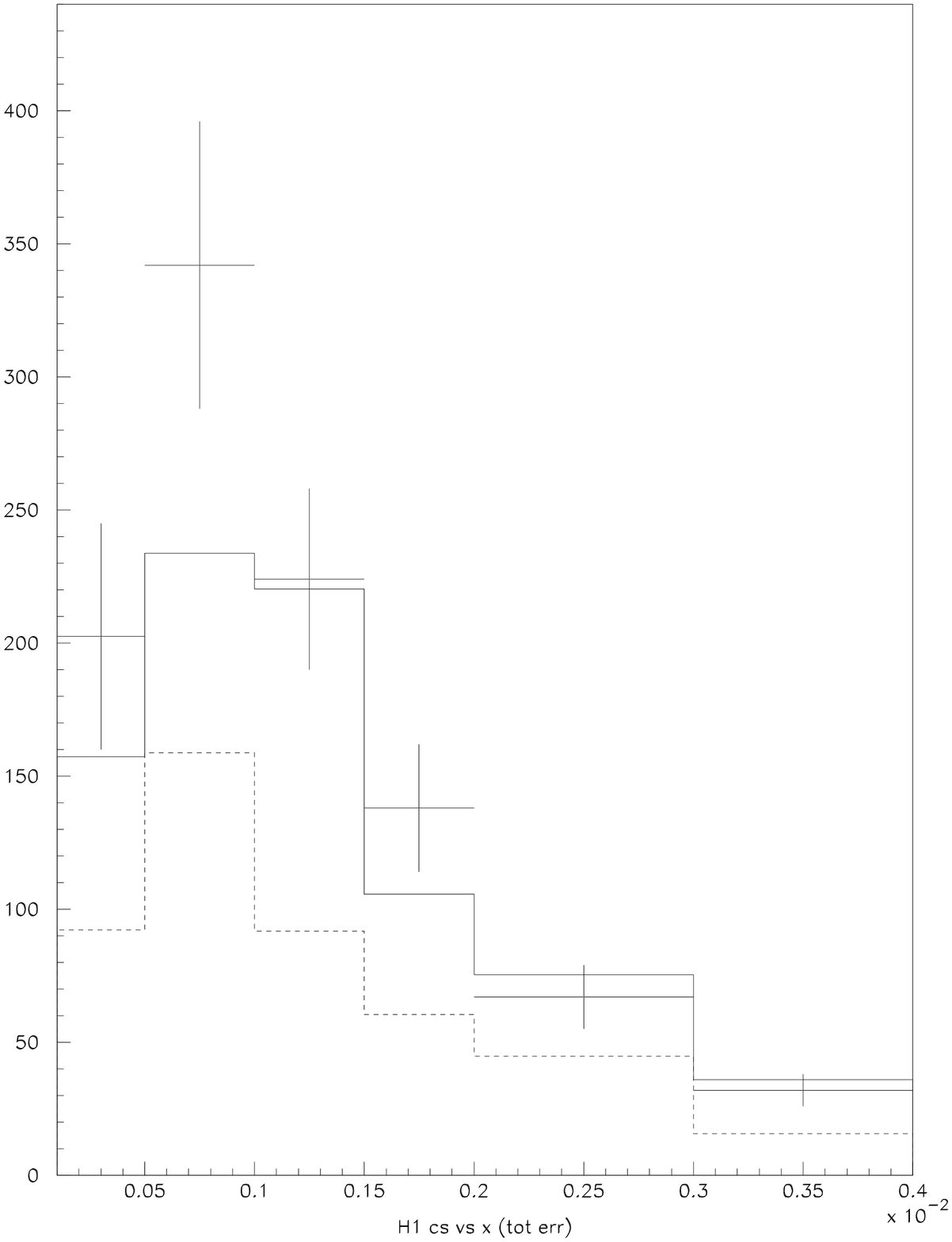,width=3.8cm,height=4.7cm}\vspace*{4pt}}
      }
    \vskip -1.3cm 
    \hspace*{-2cm}(a)\hspace*{4cm}(b)\hspace*{4cm}(c)
    \vskip 0.5cm 
  \end{center}
  \caption{
    (a) The cross section for forward jet production as a function of
    $x$, compared to H1 data~\protect\cite{H1_fjets_data}. (b) The
    values of $z$ in the initial state cascade for events satisfying
    the selection of forward jet production.  The solid (dashed) lines
    are the prediction of the Monte Carlo without (with) applying the
    ``consistency constraint'' (c.c.). (c) is the same as (a) but
    using the \protect\ldcmc\ program including only gluonic ladders
    with (dotted line) and without (full line) the non-singular terms
    in the gluon splitting function.}
  \label{fwd_jets_1}
\end{figure}
In Fig.~\ref{fwd_jets_1}a the prediction for the forward jets is
shown. The data are nicely described.  Including the consistency
constraint, the $x$ dependence of the total cross section changes, but
a similarly good description of $F_2$ is obtained by changing the
infrared cutoff, $Q_0$.  However the forward jet cross section is
reduced.

The effect of the consistency constraint is well illustrated in the
spectrum of the values of the splitting variable $z$ for events that
satisfy the criteria of forward jet production
(Fig.~\ref{fwd_jets_1}b).  Only about half of the events satisfy the
consistency constraint, and especially medium and large values of
$z$ are rejected. Furthermore we notice that in general $z$ is not
very small thus showing that even in forward jet production we are far
away from the asymptotic region, where the small $x$ approximation is
valid.

The \cascade\ program which implements CCFM evolution in a backward
evolution algorithm, gives results which agree well with the ones from
\smallx. It should also be noted that these programs also are able to
reproduce other data, such as $F_2^c$ and photo-production of $J/\Psi$
as measured at HERA.

\section{The Linked Dipole Chain Model and the LDCMC program}

The Linked Dipole Chain Model\cite{LDC}, LDC, is a reformulation of
the CCFM evolution. The main idea is to reinterpret the non-Sudakov
form factor as a normal Sudakov for the no-emission probability in
some phase-space region. By redefining the division between initial-
and final-state splittings (which is done by angular ordering in
CCFM), requiring the transverse momentum of a gluon emitted in the
initial-state to be larger than the minimum transverse momenta of the
connecting propagators, $q_{i\perp}^2 >
\min(k_{i\perp}^2,k_{i-1\perp}^2)$, an extra weight is given to each
splitting corresponding to the sum of all emissions which in this way
are treated as final-state.  In the limit of emissions which are
strongly ordered both in longitudinal and transverse momenta of the
propagating gluon, it can then be shown that if the kinematical
constraint is applied, this sum exactly cancels the non-Sudakov. It
has, however, not been proven that this holds if the consistency
constraint is relaxed.

\ldcmc\ is able to give a good description of $F_2$, but
underestimates the forward jet rates by approximately a factor 2.
Much effort has lately been put into the understanding of the
differences between \smallx\ and \cascade\ on one hand and \ldcmc\ on
the other. Some of this work was reported on in \cite{Durham}.  One
difference between \ldcmc\ and \smallx\ is that the latter only
includes the singular parts of the gluon splitting function, i.e.\ the
$1/z$ and the $1/(1-z)$ terms, where the $1/z$ term is regularized by
the non-Sudakov:
\begin{equation}
  \tilde{P}_{g}(z)= \Delta_{ns}\frac{1}{z}+\frac{1}{1-z}.
  \label{eq:split}
\end{equation}
\ldcmc, however, uses the full splitting function including the
non-singular terms (where the non-Sudakov has been canceled):
\begin{equation}
  P_{g}(z)=\frac{1}{z}+\frac{1}{1-z}-2+z(1-z).
  \label{eq:pgg}
\end{equation}
In the limits of $z$ close to zero or one, which is the limit in which
the CCFM equation is derived, the non-singular terms should be
negligible. But, as seen in fig.~\ref{fwd_jets_1}b, the typical
$z$-values in events with forward jets are around 0.5 where the
non-singular terms may reduce the splitting function with almost a
factor 2. It is intriguing to note that the forward jet rates
obtained by \ldcmc\ is approximately a factor two below those obtained
with \smallx\ (which agrees with data). Indeed, preliminary
investigations show that removing the non-singular terms from \ldcmc\ 
does increase the forward jet rates as seen in fig.~\ref{fwd_jets_1}c
although not with as much as a factor of two \cite{LeifHannes}.

\section{Open questions}

The fact that the results from the \smallx, \cascade\ and \ldcmc\ 
programs depends strongly on the treatment of the consistency
constraint and on the non-singular terms in the gluon splitting
functions, means that also the CCFM equation as such is sensitive to
these non-leading effects for observables such as the forward jet
rates.

To include the non-singular terms in the gluon splitting function in
the CCFM equation is not completely straight forward. Simply adding
them to eq.~\ref{eq:split} may result in negative splitting
probabilities. But in the limits of $z$ close to zero or one, it is
allowed to let the non-Sudakov in eq.~\ref{eq:split} multiply also the
$1/(1-z)$ term: $\tilde{P}_g= \Delta_{ns} P_{g}^{sing}(z) =
\Delta_{ns} (1/z+1/(1-z))$, in which case it is possible to replace
the $P_{g}^{sing}(z)$ with the full splitting function of
eq.~\ref{eq:pgg}. Preliminary investigations \cite{LeifHannes} show
that doing this in the \smallx\ program indeed gives a large effect,
although more studies are needed to quantify the influence on the
results for e.g.\ forward jet rates.

In conclusion, it is still an open question whether the CCFM evolution
equation is an appropriate way of describing small-$x$ final states or
not. Much more work is needed, both phenomenological work using the
event generators and purely theoretical studies of the CCFM evolution,
to understand how to correctly treat the non-leading effects which
seem to be important in the description of data.

\end{document}